\documentclass[longauth]{aa}

\usepackage{txfonts}
\usepackage{graphicx}
\usepackage{natbib}
\bibpunct{(}{)}{;}{a}{}{,}

\newcommand{\micron}{$\mu m$}

\begin{document}

\title{VVDS-SWIRE \thanks{Based on data obtained with the European Southern
    Observatory Very Large Telescope, Paranal, Chile, program 070.A-9007(A),
    and on data obtained at the Canada-France-Hawaii Telescope, operated by
    the CNRS of France, CNRC in Canada and the University of Hawaii, and
    observations obtained with MegaPrime/MegaCam, a joint project of CFHT and
    CEA/DAPNIA, at the Canada-France-Hawaii Telescope (CFHT) which is operated
    by the National Research Council (NRC) of Canada, the Institut National
    des Science de l'Univers of the Centre National de la Recherche
    Scientifique (CNRS) of France, and the University of Hawaii. This work is
    based in part on data products produced at TERAPIX and the Canadian
    Astronomy Data Centre as part of the Canada-France-Hawaii Telescope Legacy
    Survey, a collaborative project of NRC and CNRS.}}

\subtitle{Clustering evolution from a spectroscopic sample of galaxies with
  redshift $0.2<z<2.1$ selected from Spitzer IRAC 3.6\micron~and
  4.5\micron~photometry}

\titlerunning{Clustering evolution up to $z\simeq2$ of the VVDS-SWIRE spectroscopic
  sample of near-infrared galaxies}

\author{S. de la Torre \inst{1}
   \and O. Le F\`evre \inst{1}
   \and S. Arnouts \inst{1}
   \and L. Guzzo \inst{2}
   \and D. Farrah \inst{3}
   \and A. Iovino \inst{2}
   \and C.J. Lonsdale \inst{4,5}
   \and B. Meneux \inst{2,6}
   \and S.J. Oliver \inst{7}
   \and A. Pollo \inst{1,21}
   \and I. Waddington \inst{7}
   \and D. Bottini \inst{6}
   \and F. Fang \inst{10} 
   \and B. Garilli \inst{6}
   \and V. Le Brun \inst{1}
   \and D. Maccagni \inst{6}
   \and J.P. Picat \inst{13}
   \and R. Scaramella \inst{9,18}
   \and M. Scodeggio \inst{6}
   \and D. Shupe \inst{4}
   \and J. Surace \inst{10}
   \and L. Tresse \inst{1}
   \and G. Vettolani \inst{9}
   \and A. Zanichelli \inst{9}
   \and C. Adami \inst{1}
   \and S. Bardelli  \inst{8}
   \and M. Bolzonella  \inst{8} 
   \and A. Cappi    \inst{8}
   \and S. Charlot \inst{14,15}
   \and P. Ciliegi \inst{8}  
   \and T. Contini \inst{13}
   \and S. Foucaud \inst{26}
   \and P. Franzetti \inst{6}
   \and I. Gavignaud \inst{17}
   \and O. Ilbert \inst{25}  
   \and F. Lamareille \inst{12}
   \and H.J. McCracken \inst{15,16}
   \and B. Marano   \inst{12}  
   \and C. Marinoni \inst{23}
   \and A. Mazure \inst{1}
   \and R. Merighi \inst{8} 
   \and S. Paltani \inst{20,21}
   \and R. Pell\`o \inst{13}
   \and L. Pozzetti \inst{8} 
   \and M. Radovich \inst{11}
   \and G. Zamorani \inst{8} 
   \and E. Zucca    \inst{8}
   \and M. Bondi \inst{9}
   \and A. Bongiorno \inst{12}
   \and J. Brinchmann \inst{24}
   \and O. Cucciati \inst{2,19}
   \and Y. Mellier \inst{15,16}
   \and P. Merluzzi \inst{11}
   \and S. Temporin \inst{2}
   \and D. Vergani \inst{6}
   \and C.J. Walcher \inst{1}}

\offprints{\mbox{S.~de~la~Torre}, \email{sylvain.delatorre@oamp.fr}}

\institute{
Laboratoire d'Astrophysique de Marseille, UMR 6110 CNRS-Universit\'e de
Provence, BP8, 13376 Marseille Cedex 12, France
\and
INAF-Osservatorio Astronomico di Brera - Via Brera 28, Milan,
Italy
\and
Department of Astronomy, Cornell University, Space Sciences Building,
Ithaca, NY 14853, USA
\and
Infrared Processing \& Analysis Center, California Institute of
Technology, MS 100-22, Pasadena, CA 91125, USA 
\and
Center for Astrophysics \& Space Sciences, University of California San
Diego, La Jolla, CA 92093-0424, USA
\and
IASF-INAF - via Bassini 15, I-20133, Milano, Italy
\and 
Astronomy Centre, University of Sussex, Brighton BN1 9QH, UK
\and
INAF-Osservatorio Astronomico di Bologna - Via Ranzani,1, I-40127, Bologna, 
Italy
\and
IRA-INAF - Via Gobetti,101, I-40129, Bologna, Italy
\and
Spitzer Science Center, California Institute of Technology, Mail Stop 314-6, 
1200 East California Boulevard, Pasadena, CA 91125, USA
\and
INAF-Osservatorio Astronomico di Capodimonte - Via Moiariello 16, I-80131,
Napoli, Italy
\and
Universit\`a di Bologna, Dipartimento di Astronomia - Via Ranzani,1,
I-40127, Bologna, Italy
\and
Laboratoire d'Astrophysique de l'Observatoire Midi-Pyr\'en\'ees (UMR 
5572) - 14, avenue E. Belin, F31400 Toulouse, France
\and
Max Planck Institut fur Astrophysik, 85741, Garching, Germany
\and
Institut d'Astrophysique de Paris, UMR 7095, 98 bis Bvd Arago, 75014
Paris, France
\and
Observatoire de Paris, LERMA, 61 Avenue de l'Observatoire, 75014 Paris, 
France
\and
Astrophysical Institute Potsdam, An der Sternwarte 16, D-14482
Potsdam, Germany
\and
INAF-Osservatorio Astronomico di Roma - Via di Frascati 33,
I-00040, Monte Porzio Catone,
Italy
\and
Universit\'a di Milano-Bicocca, Dipartimento di Fisica - 
Piazza delle Scienze, 3, I-20126 Milano, Italy
\and
Integral Science Data Centre, ch. d'\'Ecogia 16, CH-1290 Versoix
\and
Geneva Observatory, ch. des Maillettes 51, CH-1290 Sauverny, Switzerland
\and
Astronomical Observatory of the Jagiellonian University, ul Orla 171, 
30-244 Krak{\'o}w, Poland
\and
Centre de Physique Th\'eorique, UMR 6207 CNRS-Universit\'e de Provence, 
F-13288 Marseille France
\and
Centro de Astrofísica da Universidade do Porto, Rua das Estrelas,
4150-762 Porto, Portugal 
\and
Institute for Astronomy, 2680 Woodlawn Dr., University of Hawaii,
Honolulu, Hawaii, 96822
\and    
School of Physics \& Astronomy, University of Nottingham, University
Park, Nottingham, NG72RD, UK
}

\date{Received ... / Accepted ...}

\abstract
{}
{By combining the VIMOS VLT Deep Survey (VVDS) with the Spitzer Wide-area
  InfraRed Extragalactic survey (SWIRE) data, we have built the currently
  largest spectroscopic sample of high redshift galaxies selected in the rest-frame
  near-infrared. In particular, we have obtained 2040 spectroscopic redshifts
  for a sample of galaxies with a magnitude measured at
  3.6\micron~$(m_{3.6})_{AB}<21.5$, and 1255 spectroscopic redshifts for a
  sample of galaxies with $(m_{4.5})_{AB}<21$.  These allow us to investigate,
  for the first time using spectroscopic redshifts, the clustering evolution
  of galaxies selected from their rest-frame near-infrared luminosity, in the
  redshift range $0.2<z<2.1$.}
{We use the projected two-point correlation function $w_p(r_p)$ to study the
  three dimensional clustering properties of galaxies detected at 3.6\micron
  ~and 4.5\micron~with the InfraRed Array Camera (IRAC) in the SWIRE survey
  with measured spectroscopic redshifts from the first epoch VVDS. In
  addition, we measure the clustering properties of a larger sample of 16672
  SWIRE galaxies for which we have accurate photometric redshifts on the same
  field by computing the angular correlation function. We compare these
  measurements.}
{We find that in the flux limited samples at 3.6\micron~and 4.5\micron, the
  apparent correlation length does not change from redshift $\sim2$ to the
  present.  The measured correlation lengths have a mean value of $r_0\simeq
  3.9\pm0.5~h^{-1}~Mpc$ for the galaxies selected at 3.6\micron~and a mean
  value of $r_0\simeq 4.4\pm0.5~h^{-1}~Mpc$ for the galaxies selected at
  4.5\micron, all across the redshift range explored. These values are larger
  than those typicaly found for I-band selected galaxies at $I_{AB}<24$, for
  which $r_0$ varies from $2.69~h^{-1}~Mpc$ to $3.63~h^{-1}~Mpc$ between
  $z=0.5$ to $z=2.1$. We find that the difference in correlation length
  between I-band and $3.6-4.5\mu$m selected samples is decreasing with
  increasing redshift to become comparable at $z\simeq1.5$.  We interpret this
  as evidence that galaxies with older stellar populations and galaxies
  actively forming stars reside in comparably over-dense environments at epochs
  earlier than $z\simeq1.5$, supporting the recently reported flattening of
  the color-density relation at high redshift. The increasing difference with
  cosmic time in correlation length observed between rest-frame UV-optical and
  near-infrared selected samples could then be an indication that star formation is
  gradually shifting to lower density regions as cosmic time increases, while
  the older passively evolving galaxies remain to trace the location of the
  highest primordial peaks.}
{}


\maketitle

\section{Introduction} 
According to the current cosmological paradigm, the formation of large scale
structures can be described as the evolution of primordial dark matter mass
density perturbations under the influence of gravity. The galaxies that we
observe are the result of the cooling and fragmentation of gas within the
potential wells provided by the dark matter halos, that hierarchically
build-up \citep{white78}. Even if the distribution of galaxies is in some way
biased with respect to the distribution of the dark matter \citep{kaiser84},
the spatial distribution of galaxies should therefore trace the dark matter
density field. It is to be expected that the physical processes that build
galaxies of different types and luminosities are sensitive to the mass of
halos and their different environments. Therefore the evolution of the
clustering of galaxies may depend on galaxy type, luminosity and environment.
The comparison of the clustering of different galaxy populations along cosmic
time could therefore provide useful constraints on the evolution and formation
scenario of galaxies.\\
When galaxies are observed from redder wavelengths and up to a few microns
rest-frame, this preferentially selects older stellar populations formed
earlier in the life of the Universe. As the oldest stars are likely to have
formed in the highest density peaks in the Universe, it is therefore expected
that the clustering of their host galaxies would be significantly higher than
that of galaxies hosting more recently formed stars. In the local Universe,
selecting galaxies at 2.15 microns ($K_s$ band) from the Two Micron All Sky
Survey \citep[2MASS,][]{maller05}, it is found that these galaxies are more
clustered than those selected at optical wavelengths, with an angular
correlation amplitude several times larger than found in the Sloan Digital Sky
Survey Early Data Release \citep[SDSS EDR,][]{connolly02}, even if part of
this difference may be due to the different luminosities of the samples.  At
higher redshifts, \citet{oliver04}, based on Spitzer data, found that at
$z_{median}\simeq0.75$ the correlation length of $(m_{3.6})_{AB}<20.1$ (or
equivalently $K_{Vega}<18.7$) selected galaxies is $r_0=4.4\pm0.1~h^{-1}~Mpc$
with an angular correlation amplitude $\sim60$ times lower than for 2MASS
local galaxies, and the more luminous galaxies are even more strongly
clustered \citep{farrah06}.\\
The IRAC 3.6\micron~and 4.5\micron~ bands are only marginally affected by the
first polycyclic aromatic hydrocarbon (PAH) spectral feature at 3.3\micron,
which fall in these bands in the redshift ranges [0,0.18] and [0.21,0.51]
respectively. However, galaxy populations observed in these bands contain a
mix of early-type, dust-reddened, and star-forming systems \citep{rowan05}
which make clustering expectations less straightforward. \\
This paper aims to characterize the evolution of the real-space correlation
length $r_0$ of near-infrared selected galaxies from the Spitzer Wide-area
InfraRed Extragalactic survey \citep[SWIRE,][]{swire03}, which have secure
spectroscopic redshifts from the VIMOS VLT Deep Survey
\citep[VVDS,][]{vvdsdeep05}. We compare these measurements to those in
shallower samples, e.g. \citet{oliver04}, as well as to the observed evolution
of optically-selected galaxies as shown in \citet{lefevre05}, up to $z\simeq2$.\\
In Section 2 we describe the galaxy sample that we use for this analysis. We
present in Section 3 the formalism of the two-point correlation function and
the results. Finally in Section 4 we discuss the observed clustering evolution
and we conclude.\\
Throughout this analysis we assume a flat $\Lambda CDM$ cosmology with
$\Omega_M=0.3$, $\Omega_\Lambda=0.7$, $\sigma_8=0.9$ and $H_0=100~h~km \cdot
s^{-1} \cdot Mpc^{-1}$, while a value of $H_0=70~km \cdot s^{-1} \cdot
Mpc^{-1}$ is used when computing absolute magnitudes.

\section{The VVDS-SWIRE sample}
The VVDS-SWIRE field is the intersection of the VIRMOS Deep Imaging Survey
\citep[VDIS,][]{lefevre04} and the SWIRE-XMM/LSS fields, the total area
covered on the sky is $\sim0.82~deg^2$. From this field we extract two galaxy
samples. The largest one consists of those SWIRE galaxies present in the field
for which we measured photometric redshifts, called in the following 
the \emph{photometric redshift sample}. The technique that we use to measure
photometric redshifts is presented in the next section. The other sample,
hereafter the \emph{spectroscopic redshift sample}, contains galaxies included in a
sub-area of $\sim0.42~deg^2$ which corresponds to a fraction of the VVDS-Deep
survey \citep{vvdsdeep05} area and for which we have spectroscopic redshifts
measured during the VVDS first epoch observations. We restrict this sample to
galaxies with secure redshifts, i.e. with a redshift confidence level greater
than $80\%$ \citep[flag 2 to 9,][]{vvdsdeep05}. About 25\% of the galaxy
population is randomly sampled and we are correcting the correlation
measurements for geometrical effects, e.g. on small scales, as described in
Section 3 and in \citet{pollo05}. In these two samples the star-galaxy
separation has been performed first by using optical and mid-infrared colors
to define the locus of galactic stars.  In addition we used the spectral
energy distribution (SED) template fitting procedure to reject any remaining
stars and QSO. Objects which are best fitted with a star or a QSO SED
template and which have a high SExtractor \citep{bertin96} stellarity index in
the i' band (CLASS\_STAR $>0.97$) have been removed from the samples. The full
description of the data-set and its properties is given in Le F\`evre et al.
(2007, in preparation).
\subsection{Photometric redshifts}
The photometric redshifts have been estimated using the $\chi^2$ fitting
algorithm \emph{Le Phare}\footnote{http://www.oamp.fr/arnouts/LE\_PHARE.html}
and calibrated with the VVDS first epoch spectroscopic redshifts on the same
data-set as described in \citet{ilbert06}, including in addition the
3.6\micron~and 4.5\micron~infrared photometry from SWIRE \citep{swire03}. The
photometric data cover the wavelength domain $0.3\mu m\le\lambda\le5.0\mu m$
including B,V,R,I from VDIS \citep{lefevre04}, u*,g',r',i',z' from the Canada
France Hawaii Telescope Legacy Survey (CFHTLS-D1, McCracken et al., in
preparation), K-band from VDIS \citep{iovino05} and the 3.6\micron~and
4.5\micron~infrared photometry from SWIRE \citep{swire03}.  The empirical
templates used by \citet{ilbert06} have been extended to the mid-infrared
domain using the GISSEL library \citep{bruzual03} and we use the spectroscopic
sample to simultaneously adapt the templates and the zero-point calibrations
\citep{ilbert06}. The comparison with 1500 spectroscopic redshifts from the
VVDS first epoch data shows an accuracy of $\sigma_{\Delta z/(1+z)}\simeq
0.031$ with a small fraction of catastrophic errors: $1.7\%$ objects with
$|\Delta z|>0.15(1+z)$, up to a redshift $z\sim2$.  The full description of
the method and the data used to compute these photometric redshifts is given
in \citet{ilbert06} and Arnouts et al.  (2007, in preparation).
\subsection{Selection functions}
\begin{figure}
  \resizebox{\hsize}{!}{\includegraphics{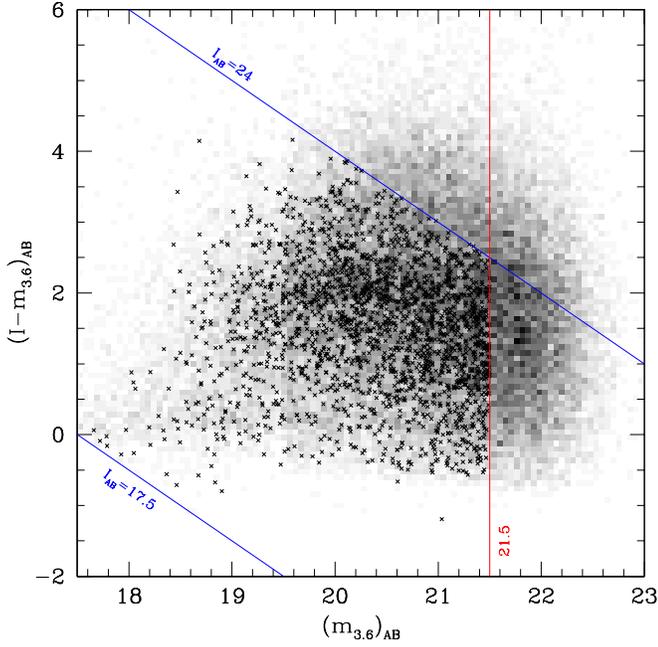}}
  \caption{Color-magnitude diagram for the 3.6\micron~spectroscopic and
    photometric sub-samples. All the galaxies detected at 3.6\micron~in the
    VVDS-SWIRE field are represented in grey scale. The photometric redshift
    sub-sample consists of all the objects with $(m_{3.6})_{AB}<21.5$ and
    galaxies from the spectroscopic sub-sample are shown as crosses. The
    vertical line indicates the limit of completeness and the two diagonal
    lines show the limits of the spectroscopic selection $17.5 < I_{AB} < 24$.
    We estimate that the fraction of galaxies we lose in the spectroscopic
    redshift sub-sample due to the VVDS spectroscopic selection and assuming
    the completeness limit $(m_{3.6})_{AB}<21.5$, is 11.6\%.}
  \label{incomp} 
\end{figure}

\begin{figure}
  \resizebox{\hsize}{!}{\includegraphics{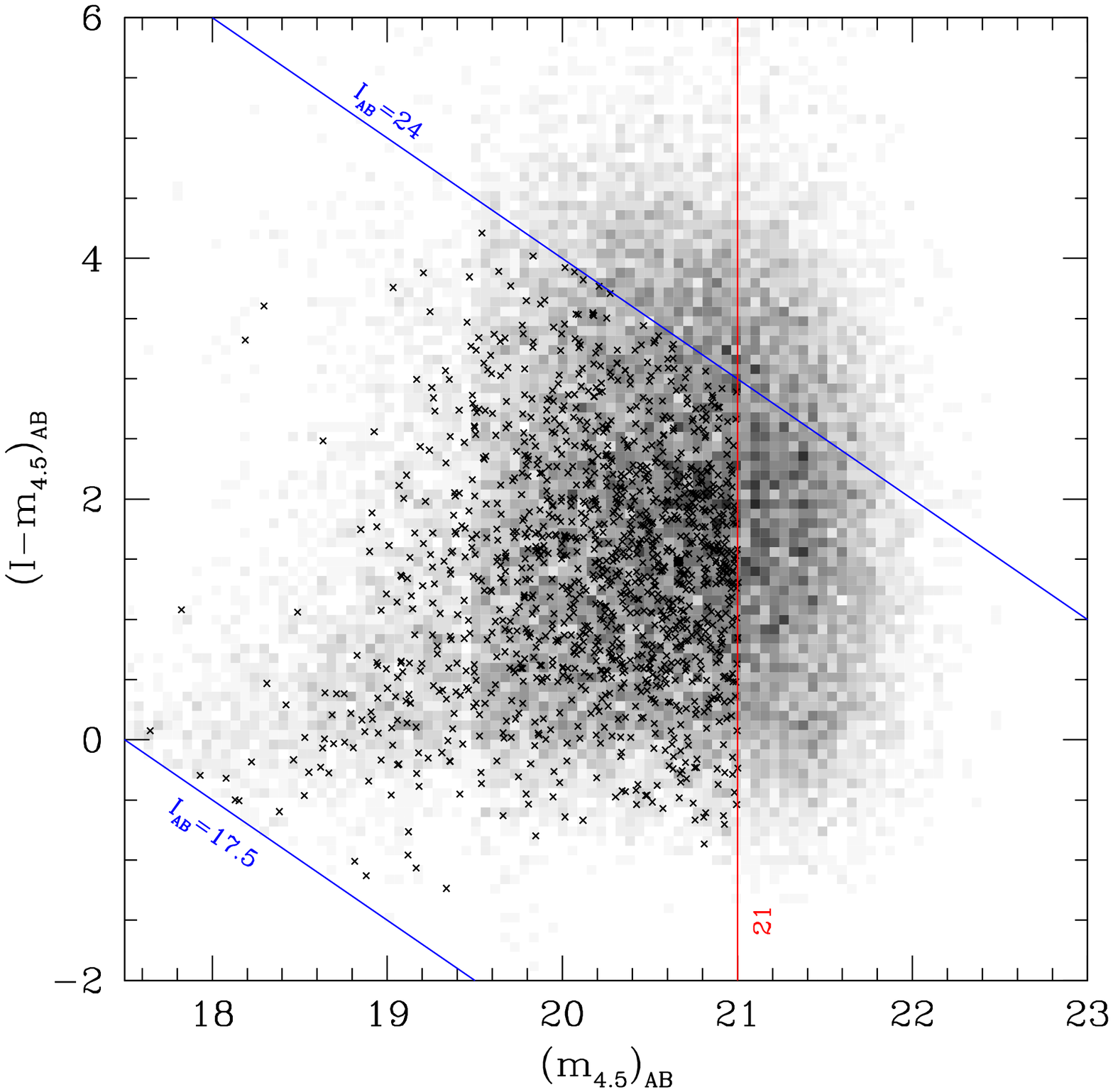}}
  \caption{Color-magnitude diagram for the 4.5\micron~spectroscopic and
    photometric sub-samples. All the galaxies detected at 4.5\micron~in the
    VVDS-SWIRE field are represented in grey scale. The photometric redshift
    sub-sample consists of all the objects with $(m_{4.5})_{AB}<21$ and
    galaxies from the spectroscopic sub-sample are shown as crosses. The
    vertical line indicates the limit of completeness and the two diagonal
    lines show the limits of the spectroscopic selection $17.5 < I_{AB} < 24$.
    We estimate that the fraction of galaxies we lose in the spectroscopic
    redshift sub-sample due to the VVDS spectroscopic selection and assuming
    the completeness limit $(m_{4.5})_{AB}<21$, is 10.3\%.}
  \label{incomp2} 
\end{figure}

\begin{figure}
  \resizebox{\hsize}{!}{\includegraphics{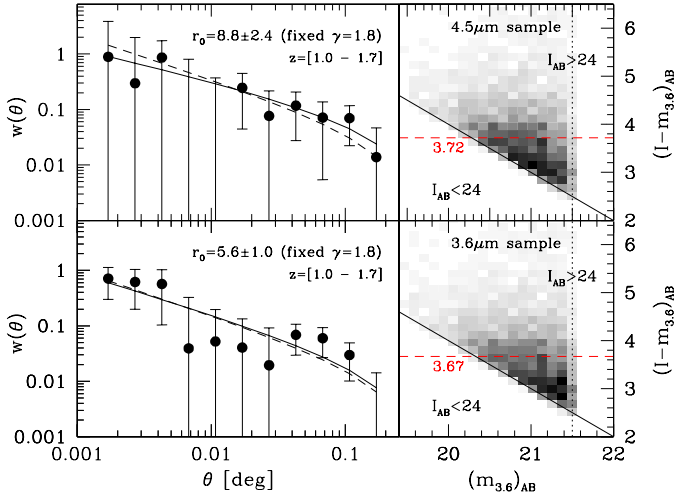}}
  \caption{The right panels present the color-magnitude diagrams
    $(I-m_{3.6})_{AB}$ versus $(m_{3.6})_{AB}$ for the 3.6\micron~(top) and
    4.5\micron~(bottom) photometric sub-samples considering the objects that
    fall above the spectroscopic selection upper limit $I_{AB}=24$.  The
    galaxy number density is shown in grey scale and the dashed line indicates
    the median of the $(I-m_{3.6})_{AB}$ distributions. In the left panels are
    shown the angular correlation function for these two selections in the
    redshift range $z_{phot}=[1,1.7]$. In these panels the solid lines show the
    best-fit model when letting the two parameters free, while the dashed
    lines indicate the best-fit model with fixed $\gamma=1.8$.}
  \label{redcolmag} 
\end{figure}

We split the spectroscopic redshift and photometric redshift samples into two
sub-samples, applying a simple magnitude selection from observed fluxes on the
two first infrared photometric bands of the SWIRE survey, the Spitzer-IRAC
3.6\micron~and 4.5\micron~bands. These sub-samples satisfy respectively to a
magnitude in the AB system measured in the 3.6\micron~band of
$(m_{3.6})_{AB}<21.5$ and in the 4.5\micron~band of $(m_{4.5})_{AB}<21$ or,
equivalently in flux, $S_{3.6}>9.2~\mu Jy$ and $S_{4.5}>14.5~\mu Jy$. These
criteria correspond to the limits of completeness for the SWIRE bands (Le
F\`evre et al., 2007, in preparation). In addition, the spectroscopic redshift
sample has a $I_{AB}$ cut due to the fact that the VVDS-Deep spectroscopic
sample is a magnitude-selected sample according to the criterion $17.5 <
I_{AB} < 24$.  The $I_{AB}$ cut introduces a color dependent incompleteness in
terms of SWIRE magnitudes in the spectroscopic redshift sample. Indeed we
loose a small fraction of objects which correspond to faint and red objects as
shown in Figure \ref{incomp} and Figure \ref{incomp2} for the 3.6\micron~and
4.5\micron~galaxies. We estimate the fraction of lost galaxies to be
respectively $11.6 \%$ and $10.3 \%$. To understand the population missed in
the spectroscopic sample, we have used the photometric redshifts described in
the previous section. We find that the galaxies that we are missing in the
4.5\micron~spectroscopic redshift sub-sample are redder than those in the
3.6\micron~spectroscopic sub-sample.  The median $(I-m_{3.6})_{AB}$ color of
these sub-samples are $3.72$ and $3.67$ respectively, as shown in Figure
\ref{redcolmag}. We note that photometric redshifts may have difficulties in
identifying dust obscured AGN because we are lacking the redder bands like the
Spitzer-IRAC 24\micron.  However, the fraction of dust reddened AGN is
expected to be low \citep{tajer07}, and its impact on clustering properties
must be insignificant.  Therefore, the photometric redshift sample, which is
assumed to be complete and free of significant contaminants, allows us to
evaluate the effect of the color dependent magnitude incompleteness on the
clustering measurements performed on the spectroscopic redshift sample.

\section{Clustering measurements} 
\subsection{Methods}
\subsubsection{Computing the projected correlation function}
The method applied to compute the projected correlation function and to
derive the real-space correlation length $r_0$ and the correlation function
slope $\gamma$, is described extensively in \citet{pollo05}. We present in
this section the main points of the method.\\
First we evaluate the bi-dimensional two-point correlation function
$\xi(r_p,\pi)$ using the standard Landy and Szalay estimator \citep{landy93}.
This estimator of the two-point correlation function is defined by:
\begin{equation}
  \xi(r)=\frac{GG(r)-2GR(r)+RR(r)}{RR(r)} \label{LS} 
\end{equation}
where $GG(r)$, $RR(r)$ and $GR(r)$ are the normalized numbers of independent
galaxy-galaxy, galaxy-random and random-random pairs with comoving separation
between $r$ and $r + dr$. The three-dimensional galaxy space distribution
recovered from a redshift survey and its two-point correlation function
$\xi(s)$, are distorted due to the effect of galaxy peculiar velocities.
Therefore the redshift-space separation $s$ differs from the true physical
comoving separation $r$. These distortions occur only radially since peculiar
velocities affect only redshift. Thus we compute the bi-dimensional two-point
correlation $\xi(r_p,\pi)$ by splitting the separation vector $s$ into two
components: $r_p$ and $\pi$, respectively perpendicular and parallel to the
line of sight \citep{fisher94}. In this way, we separate the redshift
distortions from the true spatial correlations.\\
Then, we project $\xi(r_p,\pi)$ along the line of sight, onto the $r_p$ axis.
This allows to integrate out the dilution produced by the redshift-space
distortion field and we obtain the projected two-point correlation function,
which is defined by,
\begin{equation}
  w_p(r_p)=2\int_{0}^{\infty}\xi(r_p,\pi)d\pi=2\int_{0}^{\infty}
  \xi\left(\sqrt{r_p^2+y^2}\right)dy
\end{equation}
In practice, the upper integration limit has to be finite to avoid noise,
therefore we choose its optimal value as $20~h^{-1}~Mpc$ \citep{pollo05}.
Assuming a power-law form for $\xi(r)$, i.e.
$\xi(r)=\left(r_0/r\right)^{\gamma}$, one can write $w_p(r_p)$ as,
\begin{equation}
  w_p(r_p)=r_p\left(\frac{r_0}{r_p}\right)^{\gamma}\frac
  {\Gamma\left(\frac{1}{2}\right)\Gamma\left(\frac{\gamma-1}{2}\right)}
  {\Gamma\left(\frac{\gamma}{2}\right)}
\end{equation}
where $\Gamma$ is the Euler Gamma function.\\
We fit the $w_p(r_p)$ measurements to a power-law model (equation 3) by
minimizing the generalized $\chi^2$. The advantage of this method is that it
takes into account the fact that the different points of the correlation
function are correlated. We define the generalized $\chi^2$ in our case as
\citep{pollo05},
\begin{equation}
  \chi^2=\sum_{i=1}^{N_p}\sum_{j=1}^{N_p}(w^{mod}_p(r_i)-w^{obs}_p(r_i))C^{-1}_{ij}(w^{mod}_p(r_j)-w^{obs}_p(r_j))
\end{equation}
where $w_p^{mod}(r_i)$ ($w_p^{obs}(r_i)$) is the value of $w_p^{mod}$
($w_p^{obs}$) computed at $r_p=r_i$ with $1<i<N_p$ ($N_p$ is the number of
$r_p$ points), and $C$ is the covariance matrix of the data estimated using
bootstrap re-sampling. 100 bootstrap samples have been used in this analysis.
The error bars estimated from the covariance matrix associated to the
bootstrap samples include the statistical error only and do not take into
account the error due to cosmic variance.  Therefore, in order to have a
realistic estimation of the errors bars on our measurements, we compute
$w_p(r_p)$ on 50 mock VVDS surveys constructed using the \emph{GalICS}
simulation \citep{hatton03}. This simulation of hierarchical galaxy formation
is based on a hybrid N-body/semi-analytic model and is particularly well
adapted to simulate the properties of the high-redshift galaxies that we
observe in the VVDS \citep{blaizot05,pollo05}.  Moreover, the selection
function and the observational biases present in the VVDS dataset have been
included in these mock samples \citep{pollo05}.  Thus we fit the $w_p(rp)$
measurements from the simulations using the same method as described
previously and we estimate the confidence limits on the $r_0$ and $\gamma$
measurements looking at the $\Delta \chi^2$ distribution from the best fit
\citep{pollo05}. The error bars obtained using this method are larger but more
realistic than those from the bootstrap method. They include not only the
statistical error but also the error which comes from the cosmic variance,
which dominates here.

\subsubsection{Computing the angular correlation function}
In order to compute the angular two-point correlation function $w(\theta)$
(ACF), we use the Landy \& Szalay estimator presented in equation \ref{LS},
but in this case, simply considering angular pairs. Assuming that $w(\theta)$
is well described by a power-law model as $w(\theta)=A_{w}\theta^{1-\gamma}$,
we fit the measured ACF to this model using the generalized $\chi^2$ method
and deduce the amplitude $A_{w}$ and the slope $1-\gamma$. 
Because of the finite size of the survey, we
introduce the integral constraint (IC) correction in our fitting procedure. 
It is defined as follows \citep{roche93},
\begin{equation}
  IC=\int\!\!\!\int{\frac{w(\theta)}{\Omega^2}d\Omega_{1}d\Omega_{2}}
    =\frac{A_{w}}{\Omega^2}\int\!\!\!\int{\theta^{1-\gamma}d\Omega_{1}
     d\Omega_{2}}=A_{w}B(\gamma)
\end{equation}
with,
\begin{equation}
  B(\gamma)=\frac{1}{\Omega^2}\int\!\!\!\int{\theta^{1-\gamma}d\Omega_{1}
    d\Omega_{2}}
\end{equation}
where $\Omega$ is the area of the observed field. We compute
$B(\gamma)$ by numerically integrating this expression over the entire
data area excluding regions where there are photometric defects. At
the end our measurements are fitted with the expression:
\begin{equation}
  w_{mod}(\theta)=A_{w}\theta^{1-\gamma}-IC=A_{w}(\theta^{1-\gamma}-B(\gamma))
\end{equation}
In order to derive the real-space correlation length $r_0$ from the angular
correlation amplitude $A_w$, we use the Limber deprojection technique as
described in \citet{maglio99}. This method, which is based on the Limber
relativistic equation \citep{limber53}, permits to recover the value of $r_0$
given $A_w$ and the redshift distribution of the galaxies. In general, this
technique is not very stable and the shape of the redshift distribution that
one considers is critical, but in this study, we measure ACF in relatively
narrow redshift bins so the shape we assume for the inversion is not so
critical. We use a smoothed redshift distribution by applying a moving
window with a width of $\Delta z=0.2$ to the initial photometric redshift
distribution (Le F\`evre et al., 2007, in preparation). The width of the
smoothing window has been chosen to reflect the typical uncertainty that we
have on the photometric redshift values. The ACF error bars have been
estimated using the bootstrap re-sampling method.

\subsection{Clustering results}
\subsubsection{Clustering of the spectroscopic redshift sample}
\begin{figure*}
  \centering
  \includegraphics[width=12cm]{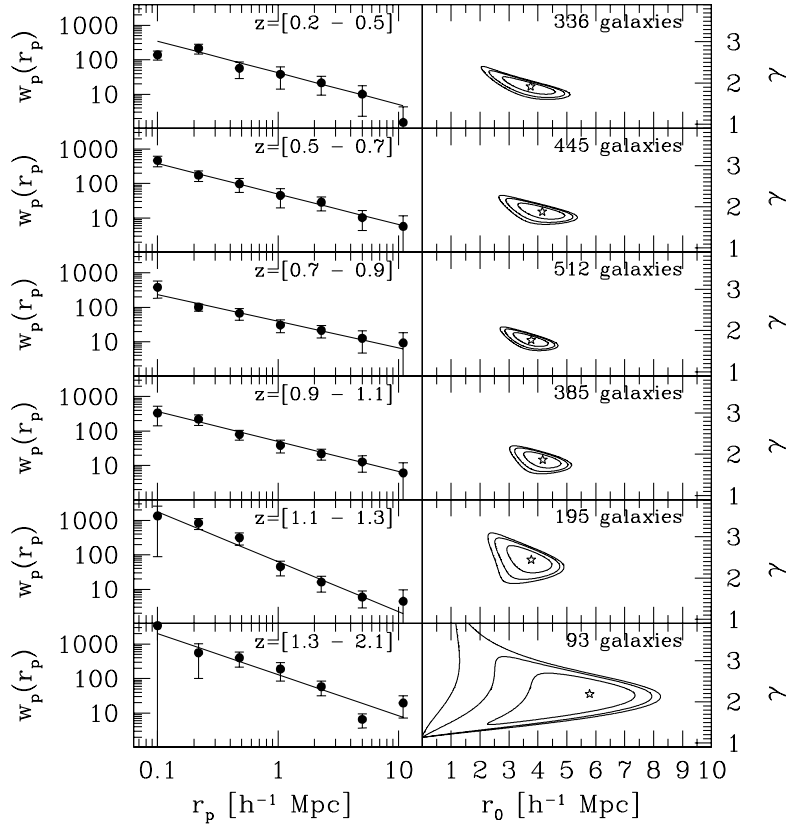}
  \caption{ILeft panels: projected correlation functions
    for the 3.6\micron~spectroscopic redshift sub-sample in different redshift
    slices with their best-fit power law. Right panels: the $68.3\%$,
    $90\%$ and $95.4\%$ confidence contours for the clustering parameters
    $r_0$ and $\gamma$.}
  \label{fig5}
\end{figure*}

We compute $w_p(r_p)$ in increasing spectroscopic redshift slices from $z=0.2$
to $z=2.1$. Therefore we divide the 3.6\micron~sub-sample in six redshift
slices: $[0.2,0.5]$, $[0.5,0.7]$, $[0.7,0.9]$, $[0.9,1.1]$, $[1.1,1.3]$,
$[1.3,2.1]$. The 4.5\micron~sub-sample has fewer objects than the
3.6\micron~one because of the slightly shallower depth. In order to maximize
the number of objects per redshift slice, we consider in this
case three slices only: $[0.2,0.7]$, $[0.7,1.1]$ and $[1.1,2.1]$.\\
In the fitting procedure, we use all the $w_p(r_p)$ points for $0.1\le r_p \le
12~h^{-1}~Mpc$. The projected correlation functions computed in each redshift
slice are plotted in Figure \ref{fig5} with the contours of confidence for the
best-fitted parameters. Table \ref{table:1} summarizes the values of the
correlation length and the correlation function slope that
we derive for the two sub-samples of the spectroscopic redshift sample.\\
\begin{table*}
  \caption{VVDS-SWIRE spectroscopic redshift sample: sub-sample properties and
    associated measurements of the slope, the correlation length and the
    linear bias.}
\label{table:1}
\centering
\begin{tabular}{ccccccccc}
  \hline\hline
  \multicolumn{2}{l}{Spectroscopic redshift sample} \\
  \hline
  Sub-sample & Redshift & Mean & Number & $\langle M_B\rangle_{Vega}$ &
  $\gamma$ & $r_0$ ($h^{-1}~Mpc$) & $r_0$ ($h^{-1}~Mpc$) & $b_L$\\
  & interval & redshift & of galaxies &&&& (fixed $\gamma=1.8$) & \\
  \hline
  3.6\micron & $[0.2,0.5]$ & 0.36 & 336 &-19.53 & $1.92^{+0.27}_{-0.18}$ &
  $3.8^{+0.6}_{-0.7}$ & $3.9^{+0.6}_{-0.7}$ & $0.93^{+0.10}_{-0.11}$\\
  & $[0.5,0.7]$ & 0.60 & 445 &-20.29 & $1.88^{+0.21}_{-0.14}$ &
  $4.2^{+0.5}_{-0.5}$ & $4.4^{+0.5}_{-0.5}$ & $1.15^{+0.10}_{-0.10}$\\
  & $[0.7,0.9]$ & 0.81 & 512 &-20.75 & $1.77^{+0.15}_{-0.11}$ &
  $3.8^{+0.4}_{-0.4}$ & $3.7^{+0.4}_{-0.4}$ & $1.15^{+0.09}_{-0.09}$\\
  & $[0.9,1.1]$ & 0.99 & 385 &-21.17 & $1.87^{+0.17}_{-0.14}$ &
  $4.2^{+0.4}_{-0.4}$ & $4.3^{+0.4}_{-0.4}$ & $1.38^{+0.11}_{-0.11}$\\
  & $[1.1,1.3]$ & 1.19 & 195 &-21.66 & $2.44^{+0.22}_{-0.16}$ &
  $3.8^{+0.5}_{-0.5}$ & $3.7^{+0.5}_{-0.5}$ & $1.62^{+0.43}_{-0.43}$\\
  & $[1.3,2.1]$ & 1.57 & 93  &-22.13 & $2.19^{+0.37}_{-0.38}$ &
  $5.8^{+0.7}_{-1.0}$ & $7.6^{+0.7}_{-1.0}$ & $2.55^{+0.76}_{-0.96}$\\
  4.5\micron & $[0.2,0.7]$ & 0.48 & 485 &-20.22 & $1.84^{+0.18}_{-0.16}$ &
  $3.9^{+0.5}_{-0.5}$ & $3.9^{+0.5}_{-0.5}$ & $1.01^{+0.08}_{-0.08}$\\
  & $[0.7,1.1]$ & 0.90 & 537 &-21.15 & $1.84^{+0.14}_{-0.14}$ &
  $4.0^{+0.3}_{-0.4}$ & $4.0^{+0.3}_{-0.4}$ & $1.26^{+0.09}_{-0.10}$\\
  & $[1.1,2.1]$ & 1.42 & 182 &-21.97 & $1.82^{+0.26}_{-0.22}$ &
  $3.7^{+0.6}_{-0.7}$ & $3.7^{+0.6}_{-0.7}$ & $1.48^{+0.24}_{-0.28}$\\
  \hline
\end{tabular}
\end{table*}
The correlation length measured in the $3.6$\micron~and
$4.5$\micron~sub-samples is roughly constant over the redshift range
z=[0.2,2.1]. It varies slightly between $3.8$ and $4.2~h^{-1}~Mpc$, except in
the high-redshift interval $[1.3,2.1]$ of the $3.6$\micron~sub-sample where a
large correlation length of $5.8~h^{-1}~Mpc$ is found. In this redshift range
we are probing the intrinsically brightest galaxies, with a median absolute
magnitude of $\langle M_B\rangle_{Vega}=-22.13$ (Table \ref{table:1}).
Compared to the characteristic absolute magnitude $M^*_B$ at this redshift
reported by \citet{ilbert05}, it is found that for these objects, $\Delta
M^*_B=\langle M_B\rangle-M_B^*\simeq-0.27$ and equivalently $\langle
L_B\rangle/L^*_B \simeq 1.3$. Thus, these luminous galaxies are expected to be
more strongly clustered as the halo clustering strength increases more rapidly
for luminosities greater than a few times $L^*$ \citep{zehavi05,norberg02}.
Indeed \citet{pollo06,coil06} have shown that the galaxy correlation length
increases more steeply for magnitude brighter than $M^*$ at $z\simeq1$ with
$r_0$ values about $5~h^{-1}~Mpc$. Our high $r_0$ measurement in this high
redshift bin is thus fully consistent with these findings.  Moreover, in this
redshift interval the 1.6\micron~rest-frame emission feature from the
photospheric emission of evolved stars \citep{simpson99}, is present in the
3.6\micron~band, favoring the detection of more massive and significantly more
clustered systems \citep{farrah06}. We discuss in details the evolution of the
correlation length for the two sub-samples in Section 4.\\

\subsubsection{Clustering of the photometric redshift sample} 
\begin{figure*}
  \centering
  \includegraphics[width=12cm]{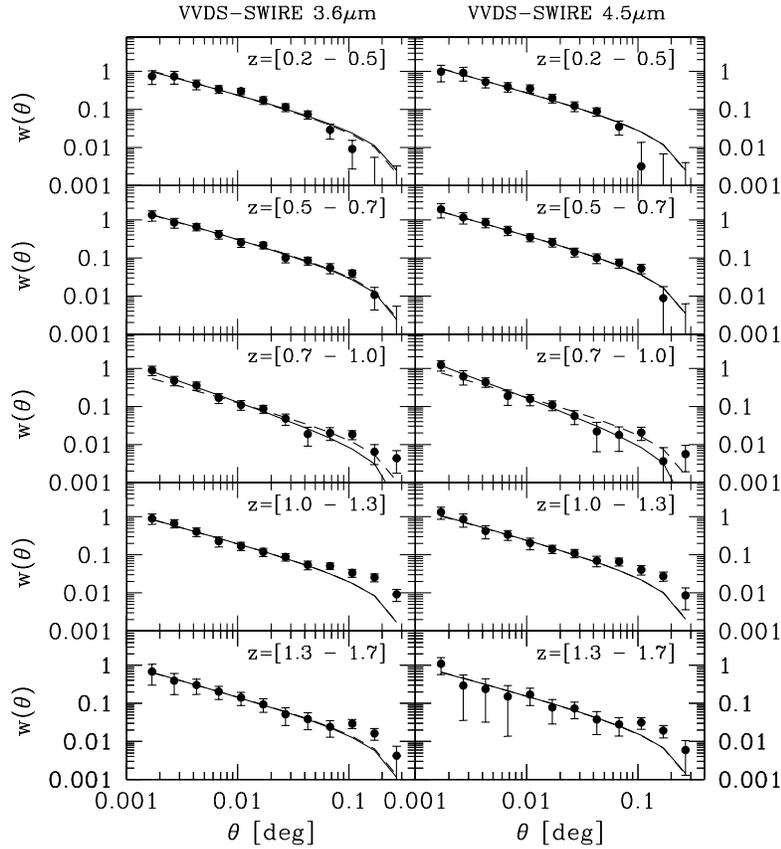}
  \caption{The angular correlation function as a
    function of redshift, for the 3.6\micron~and 4.5\micron~photometric
    redshift sub-samples.  The solid lines show the best-fit model when letting
    the two parameters free, while the dashed lines indicate the best-fit
    model with fixed $\gamma=1.8$.}
    \label{fig4}
  \end{figure*}

We measure the ACF and deduce the clustering parameters $r_0$ and $\gamma$ on
the photometric redshift sample in order to test the measurements from the
spectroscopic redshift sample and to evaluate the effect of incompleteness.\\
With the same methodology we compute the ACF in increasing photometric
redshift slices, selecting the slice boundaries in order to take the
photometric redshift errors into account. We consider five slices with
increasing size: $[0.2,0.5]$, $[0.5,0.7]$, $[0.7,1.0]$, $[1.0,1.3]$ and
$[1.3,1.7]$. No redshift slices have been considered beyond $z_{phot}=1.7$
because the lack of accuracy of the photometric redshifts in this case
seriously corrupts the measurements.\\
We fit the measured ACF to a power-law model over $0.0017\le\theta\le0.1$
degrees, where the shape of the ACF is very close to a power law.  However,
it is evident in Figure \ref{fig4} that the ACF departs from this model 
at scales greater than $\sim0.1$ degrees in redshift slices with 
$z_{phot}>0.7$. The excess of power at large angular scale could be
explained by the presence of another component in the correlation function or
of the presence of residual inhomogeneities in the field. We will explore this
point more in detail in a future paper as it does not impact the results
presented here. The ACF for the different redshift slices are
plotted in Figure \ref{fig4}. The
best-fitted values of the angular correlation amplitude $A_w$, the correlation
function slope $\gamma$ and the correlation length $r_0$ that we derive with
the Limber deprojection technique, are reported in Table \ref{table:2}.\\
\begin{table*}
\caption{VVDS-SWIRE photometric redshift sample: sub-sample properties and associated
  measurements of the ACF amplitude, the slope and the correlation length.}
\label{table:2}
\centering
\begin{tabular}{ccccccccc}
  \hline\hline
  \multicolumn{2}{l}{Photometric redshift sample} \\
  \hline
  Sub-sample & Photometric & Mean redshift & Number of galaxies & $A_w
  \times 10^3$ & $\gamma$ & $r_0$ ($h^{-1}~Mpc$) & $r_0$ ($h^{-1}~Mpc$) \\
  & redshift interval &&&&&& (fixed $\gamma=1.8$) \\  
  \hline
  3.6\micron & $[0.2,0.5]$ & 0.37 & 2726 & $7.7^{+4.7}_{-3.1}$ & $1.76^{+0.10}_{-0.09}$ & $4.2^{+1.5}_{-1.0}$ & $4.0^{+0.3}_{-0.3}$\\
  & $[0.5,0.7]$ & 0.60 & 2714 & $7.1^{+4.5}_{-3.0}$ & $1.83^{+0.10}_{-0.09}$ & $4.1^{+1.4}_{-0.9}$ & $4.2^{+0.3}_{-0.3}$\\
  & $[0.7,1.0]$ & 0.85 & 5110 & $1.2^{+0.8}_{-0.5}$ & $2.03^{+0.11}_{-0.11}$ & $2.8^{+1.0}_{-0.6}$ & $3.4^{+0.2}_{-0.2}$\\
  & $[1.0,1.3]$ & 1.14 & 3577 & $5.1^{+4.0}_{-2.5}$ & $1.81^{+0.12}_{-0.11}$ & $4.3^{+1.9}_{-1.1}$ & $4.3^{+0.3}_{-0.3}$\\
  & $[1.3,1.7]$ & 1.47 & 2545 & $3.4^{+3.7}_{-1.9}$ & $1.83^{+0.16}_{-0.15}$ & $4.0^{+2.4}_{-1.2}$ & $4.1^{+0.4}_{-0.4}$\\
  4.5\micron & $[0.2,0.5]$ & 0.37 & 1977 & $7.6^{+4.3}_{-2.8}$ & $1.79^{+0.09}_{-0.10}$ & $4.4^{+1.4}_{-0.9}$ & $4.3^{+0.3}_{-0.3}$\\
  & $[0.5,0.7]$ & 0.60 & 1708 & $10.4^{+8.5}_{-4.9}$ & $1.80^{+0.13}_{-0.12}$& $4.8^{+2.2}_{-1.2}$ & $4.8^{+0.3}_{-0.3}$\\
  & $[0.7,1.0]$ & 0.85 & 2971 & $1.1^{+1.0}_{-0.5}$ & $2.10^{+0.12}_{-0.13}$ & $3.0^{+1.4}_{-0.7}$ & $4.0^{+0.3}_{-0.3}$\\
  & $[1.0,1.3]$ & 1.14 & 2287 & $6.1^{+7.1}_{-3.3}$ & $1.81^{+0.15}_{-0.16}$ & $4.8^{+3.1}_{-1.4}$ & $4.8^{+0.4}_{-0.4}$\\
  & $[1.3,1.7]$ & 1.47 & 1700 & $4.3^{+5.9}_{-2.5}$ & $1.79^{+0.17}_{-0.19}$ & $4.3^{+3.2}_{-1.4}$ & $4.2^{+0.5}_{-0.5}$\\
  \hline
\end{tabular}
\end{table*}
  Given the uncertainties on the ACF measurements, we adopt a fixed slope
  $\gamma=1.8$.  This is supported by the fact that when letting both $r_0$
  and $\gamma$ vary in the ACF fitting, we find $\gamma$ values very similar
  to 1.8, while the errors on $r_0$ increase slightly.\\
  The values of the correlation length that we derive for the 3.6\micron~and
  4.5\micron~photometric samples do not evolve with redshift up to
  $z_{phot}=1.7$. In the case of the 3.6\micron~sub-sample, the $r_0$ values
  are in good agreement with the values found in the spectroscopic redshift
  sample given the size of the error bars. We observe slightly larger values
  of the correlation length for the 4.5\micron~photometric redshift sub-sample
  than for the 4.5\micron~spectroscopic redshift one. We interpret this as an
  effect of the incompleteness of the spectroscopic samples due to the I-band
  selection. To test this we have measured on the photometric sample the
  clustering of the red population which is missed by the spectroscopic sample
  in each of the 3.6\micron~and 4.5\micron~bands. As most of the missed
  population in both cases is at redshifts $z_{phot}>1$, we have been able to
  perform this analysis only in the redshift range $z_{phot}=[1,1.7]$ where
  the available number of galaxies is large enough for a clustering signal to
  be measured. Results are presented in Figure \ref{redcolmag}. We find that
  the clustering of the missed red population in the 3.6 microns spectroscopic
  sample has a correlation length of $r_0=5.6\pm1.0$ only marginally larger
  than that of the bulk of the population, and therefore the clustering signal
  of the photometric and spectroscopic samples are found to be very similar.
  For the red population in the 4.5 microns sample, we find that
  $r_0=8.8\pm2.4$, a significantly larger value that the main population.
  This explains why the spectroscopic sample slightly underestimates the
  clustering signal when compared to the photometric sample.

\section{Discussion and conclusions}

\begin{figure}
  \resizebox{\hsize}{!}{\includegraphics{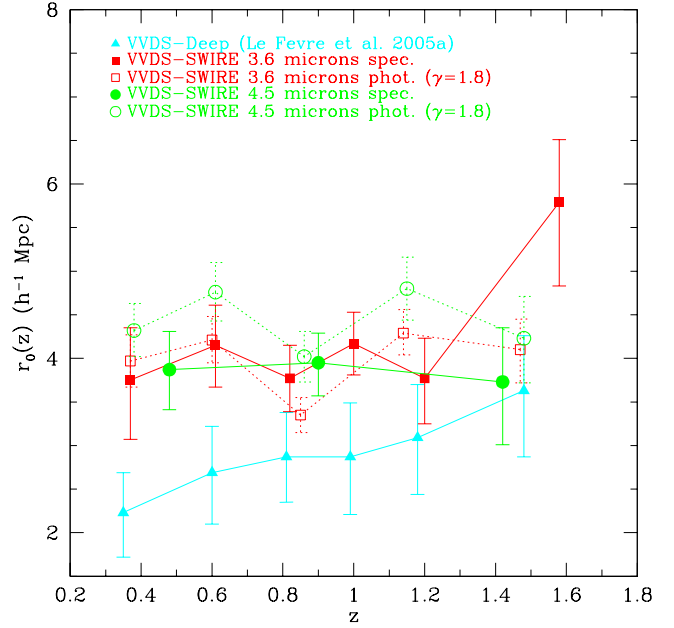}}
  \caption{Clustering evolution of VVDS-SWIRE samples of near-infrared
    galaxies compared to the VVDS-Deep I-band measurements. Filled symbols
    refer to measurements from spectroscopic samples while open symbols refer
    to measurements from photometric samples.}
  \label{fig2} 
\end{figure}
\begin{figure}
  \resizebox{\hsize}{!}{\includegraphics{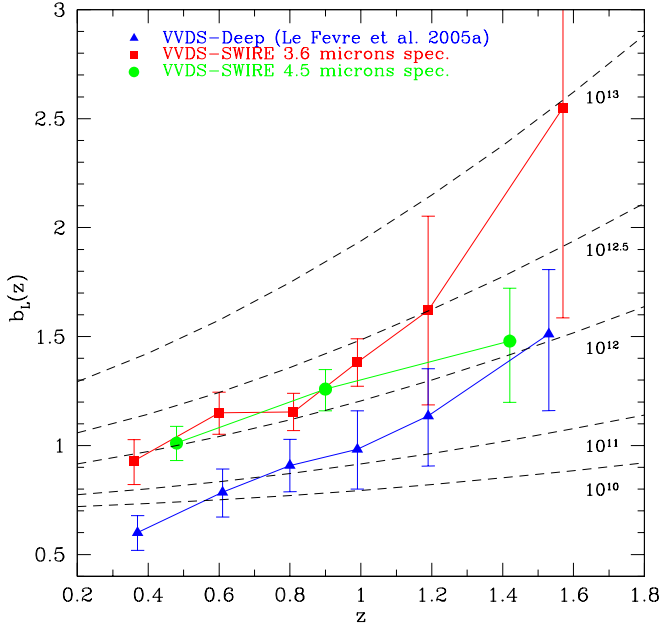}}
  \caption{Trend of the linear bias as a function of redshift for the
    VVDS-SWIRE 3.6\micron~and 4.5\micron~spectroscopic redshift samples
    compared to the VVDS-Deep I-band one. The dashed curves show the
    theoritical linear halo bias evolution for dark matter halos of mass
    greater than $10^{10},10^{11},10^{12},10^{12.5}$ and $10^{13} h^{-1}
    M_\odot$ (from bottom to top). The linear galaxy bias is computed as
    described in \citet{maglio00} from the $r_0$ and $\gamma$ measurements.
    The theoretical curves have been computed using the fitting functions of
    \citet{sheth99} for the halo mass function and bias.}
  \label{bias} 
\end{figure}
\begin{figure}
  \resizebox{\hsize}{!}{\includegraphics{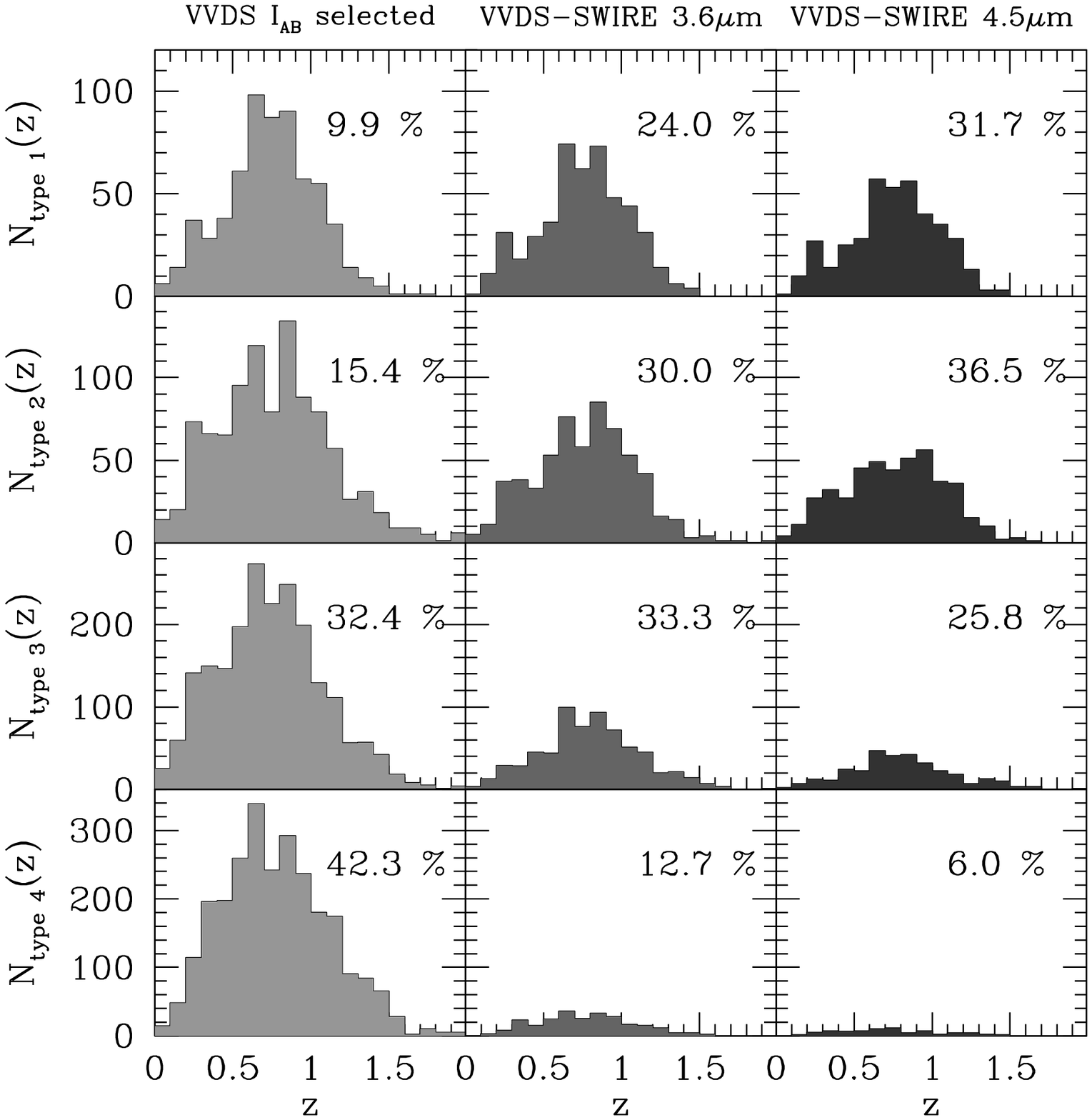}}
  \caption{Redshift distribution per spectral types for the 3.6\micron~and
    4.5\micron~spectroscopic redshift sub-samples compared to the VVDS-Deep
    sample. We use the spectral types defined in \citet{zucca06}: type 1
    corresponds to E/S0, type 2 to early spiral, type 3 to late spiral and 
    type 4 to irregular/starburst galaxies.}
  \label{fig3}
\end{figure}
We first compare our measurements with previously reported results on samples
of galaxies selected using Spitzer-IRAC bands.  \citet{oliver04} use a similar
type of galaxy selection, as they define a flux-limited sample of SWIRE
galaxies with $S_{3.6}> 32\mu Jy$, and they compute the ACF and deduce $r_0$
using Limber deprojection assuming a parametrized redshift distribution. The
selection criterion they consider is different from the one we use, i.e.
$S_{3.6}> 9.2\mu Jy$. This selects brighter galaxies than in our sample, that
are expected to be more strongly clustered \citep{pollo06,coil06}. They find a
correlation length $r_0=4.4 \pm 0.1~h^{-1}~Mpc$ at $z\simeq0.75$ with
$\gamma=1.8$, which is higher than our value of $r_0=3.7 \pm 0.4~h^{-1}~Mpc$
at $z\simeq0.81$. Taking into account that galaxies in their sample are 1.4
magnitudes brighter and that brighter galaxies are on average more clustered,
the two results are found to be in good agreement.\\
We then compare our results with the clustering measurements of I-band
selected VVDS-Deep galaxies \citep{lefevre05} in the same area. This
comparison is shown in Figure \ref{fig2}, where the correlation length for the
different samples is plotted. We observe that the correlation length is
globally higher for near-infrared selected galaxies compared to
optically-selected ones and that it is roughly constant over the redshift
range $z=[0.2,2.1]$.  By comparison, $r_0$ from the I-selected sample shows a
steady increase with redshift. As discussed in Section 3, the clustering of
the 4.5\micron-selected galaxy population is higher as measured from the
photometric sample than from the spectroscopic sample because the latter
misses a significant fraction of red galaxies due to the additional $17.5 <
I_{AB} < 24$ cut applied by the VVDS-Deep survey selection. When comparing the
measurements at 3.6\micron~ and 4.5\micron~ obtained from the photometric
sample (open symbols in Figure 6), we observe that the clustering of the
sample selected at the reddest wavelength is slightly stronger, although
comparable given the errors. This is expected, as selecting from longer
wavelengths we are sampling preferentially more early-types than later ones as
described below. The color dependency of the clustering of galaxies in the
VVDS-SWIRE sample will be reported elsewhere (\mbox{de~la~Torre} et al.,
in preparation).\\
In order to understand the systematically higher values of clustering that we
find compared to those from optically-selected galaxies and to quantify the
type of galaxy population included in our near-infrared samples, we derive the
spectral types of our spectroscopic redshift sample galaxies using the galaxy
classification introduced by \citet{zucca06}. This classification is based on
the match to an empirical set of spectral energy distributions \citep[as
described in][]{arnouts99} of the galaxy rest-frame colors, obtained from the
multi-wavelength information and spectroscopic redshift. The redshift
distributions per galaxy types are presented in Figure \ref{fig3} for the
I-band, 3.6\micron~and 4.5\micron~selected samples. The four spectral types
defined as type 1 to 4 correspond respectively to E/S0, early spiral, late
spiral and irregular/starburst galaxies. One can see in Figure \ref{fig3} that
the VVDS-Deep I-selected sample is dominated by late types (type 3 and 4)
whereas the 4.5\micron~selected sample is dominated by early types (type 1 and
2). In the 3.6\micron~sample the different types balance out. Early type
galaxies are known to be more clustered than late type ones at $z=0$
\citep[e.g.][]{norberg02,zehavi05} and also at high redshift \citep{meneux06},
thus the proportion of early type galaxies that we found in the different
samples can explain why globally we measure higher values of clustering when
galaxies are selected from redder observed wavelengths.\\
On the other hand, understanding the relatively constant clustering as a
function of redshift that we observe in our near-infrared samples is not
straightforward.  Selecting galaxies at 3.6\micron~or 4.5\micron, we are
probing the rest-frame near-infrared of galaxies, more sensitive to the
emission from old stars, and therefore our near-infrared selection is mostly
driven by stellar mass.  As dark matter halo clustering is expected to
decrease with increasing redshift \citep{weinberg04}, and under a reasonable
assumption that stellar mass traces the underlying dark matter halo mass, one
would then expect the correlation length of our sample to decrease with
increasing redshift. On the contrary, as we select intrinsically more luminous
galaxies with increasing redshift because of the magnitude selection of the
sample, and since more luminous galaxies are more clustered
\citep{pollo06,coil06}, we expect higher clustering strength as redshift
increases. We suggest that these two competing effect, working simultaneously,
combine to produce the roughly constant clustering amplitude with redshift 
that we observe.\\
In order to check this hypothesis we look at the evolution of the linear
galaxy bias for the different samples. We compute the galaxy linear bias $b_L$
as described in \citet{maglio00} from our $r_0$ and $\gamma$ measurements. We
define $b_L$ as,
  \begin{equation}
    b_L(z)=\frac{\sqrt{C_\gamma\big(r_0(z)/8\big)^\gamma}}{\sigma_{8,mass}(z)}
    \textrm{~~,~~}
    C_\gamma=\frac{72}{(3-\gamma)(4-\gamma)(6-\gamma)2^\gamma}
  \end{equation}
  This assumes a linear evolution of the rms mass fluctuations with cosmic
  time, i.e.  $\sigma_{8,mass}(z)=\sigma_{8,mass}(z=0)D(z)$, $D(z)$ being the
  linear growth factor. We compare it to the theoretical halo bias evolution
  for dark matter halos of constant mass. To evaluate the linear halo bias we
  use the fitting functions of \citet{sheth99} for the halo mass function and
  bias. As shown in Figure \ref{bias} the linear galaxy bias evolves more
  rapidly than the linear halo bias, indicating that these galaxies are not
  strictly mass-selected. Indeed because of our magnitude selection, we are
  probing more luminous, hence more biased galaxies at increasing redshifts.
  However, at least up to $z\simeq1.5$ (and given the uncertainties), the
  linear galaxy bias for the rest-frame near-infrared galaxy samples seems to
  stay in between the linear halo bias tracks for halos of mass greater than
  $10^{12}$ and $10^{12.5}~h^{-1} M_\odot$. This spread in mass is lower than
  the one for the rest-frame UV-optical VVDS-Deep sample, making this sample
  more suitable to trace the passive growth of clustering in the mass. This
  comparison of the linear galaxy bias and the linear halo bias is consistent
  with the idea that the clustering dependence on luminosity and stellar mass
  combine to produce a relatively constant clustering with redshift as
  suggested in the previous paragraph.\\
  It is particularly interesting to note that at $z\simeq1.5$, the correlation
  length of galaxies selected from the UV-optical rest-frame and hence
  actively star-forming is very similar to that of galaxies selected in the
  rest-frame near-infrared dominated by older stellar populations. This is
  quite different from what is observed in the local Universe where selecting
  galaxies in the $K_s$ band from the 2MASS survey \citep{maller05}, it is
  found that these galaxies are significantly more clustered than those
  selected at optical wavelengths \citep[SDSS EDR,][]{connolly02}. The
  galaxies which are making stars at $z\simeq1.5$ therefore reside in
  similarly clustered regions as those containing the bulk of stellar mass,
  while at $z=0$ the star formation happens in a population which is much less
  clustered than that where the bulk of stellar mass already resides.  This
  result is consistent with the observed flattening of the color-density
  relation above $z=1.2-1.5$ \citep{cucciati06}, which indicates that at these
  redshifts, blue star forming galaxies and red older galaxies are found to
  reside with equal probability in high or low density regions. Furthermore,
  we note that the clustering of massive luminous galaxies at $z\simeq1.5$ is
  measured to be $r_0\simeq 4~h^{-1}~Mpc$, independently of the selection at
  optical or near-infrared wavelengths, while at lower redshifts, the
  difference in the clustering strength measured from these samples increases.
  Galaxies selected from their UV-optical rest-frame show lower and lower
  clustering, whereas galaxies selected in the near-infrared (i.e. more
  mass-selected) stay equally clustered.  Taken together, this can be
  interpreted as evidence for star formation shifting from high density to low
  density regions, as cosmic time increases, another manifestation of the
  \emph{downsizing} trend which has been indicated by observations over the
  last few years \citep[e.g.][]{cowie96}.  This underlines the primary role of
  the mass in regulating star formation in galaxies \citep{gavazzi02}: while
  at high redshift star formation was strong in massive and more clustered
  galaxies, today it is mainly limited to lower mass galaxies more uniformly
  distributed.
\appendix

\begin{acknowledgements}
  This research has been developed within the framework of the VVDS
  consortium.\\
  This work has been partially supported by the CNRS-INSU and its Programme
  National de Cosmologie (France), and by Italian Ministry (MIUR) grants
  COFIN2000 (MM02037133) and COFIN2003 (num.2003020150).\\
  The VLT-VIMOS observations have been carried out on guaranteed time (GTO)
  allocated by the European Southern Observatory (ESO) to the VIRMOS
  consortium, under a contractual agreement between the Centre National de la
  Recherche Scientifique of France, heading a consortium of French and Italian
  institutes, and ESO, to design, manufacture and test the VIMOS instrument.
\end{acknowledgements}

\bibliographystyle{aa}
\bibliography{clust_swire}

\end{document}